\documentclass[apl,reprint]{revtex4-1}
\usepackage{graphicx}
\usepackage{bm,braket,color,ulem}
\usepackage{amssymb}

\begin{document}

\title{Low-temperature-compatible tunneling-current-assisted scanning microwave microscope utilizing a rigid coaxial resonator}

\author{Hideyuki Takahashi$^{1,3}$, Yoshinori Imai$^{2,3}$, Atsutaka Maeda$^{3}$}
\affiliation{
$^{1}$Organization for Advanced and Integrated Research, Kobe University, 1-1, Rokkodai, Nada, Kobe 657-8501, Japan\\
$^{2}$Department of Physics, Tohoku University, 6-3, Aramaki Aza-Aoba, Aoba-ku, Sendai 980-8578, Japan\\
$^{3}$Department of Basic Science, the University of Tokyo, 3-8-1 Komaba, Meguro-ku, Tokyo 153-8902, Japan}
\date{\today}

\begin{abstract}
We present a design for a tunneling-current-assisted scanning near-field microwave microscope.
For stable operation at cryogenic temperatures, making a small and rigid microwave probe is important.
Our coaxial resonator probe has a length of approxomately 30 mm and can fit inside the 2-inch bore of a superconducting magnet.
The probe design includes an insulating joint, which separates DC and microwave signals without degrading the quality factor.
By applying the SMM to the imaging of an electrically inhomogeneous superconductor, we obtain the spatial distribution of the microwave response with a spatial resolution of approximately $200\ \mathrm{nm}$.
Furthermore, we present an analysis of our SMM probe based on a simple lumped-element circuit model along with the near-field microwave measurements of silicon wafers having different conductivities. 
\end{abstract}

\maketitle
\section{Introduction}
A scanning near-field microwave microscope (SMM) enables local electrical characterization with a spatial resolution beyond the diffraction limit.
Since the first SMM was produced by Ash and Nicholls~\cite{Ash1972}, various such instruments have been reported.
They are divided into two categories; apertured~\cite{Ash1972,TabibAzar1999,Park2001} and apertureless probes~\cite{Wei1996}.
Recent progress has mainly been on the latter type.
The spatial resolution of the apertureless-type SMM has significantly improved by precise control of the tip-sample distance.
Using the feedback control circuits of a scanning tunneling microscope (STM-SMM~\cite{Imtiaz2003,Machida2009,Lee2010})  and an atomic force microscope (AFM-SMM~\cite{Kim2003,Gao2005,TabibAzar2004,Lai2007,Zhang2010}), a nanometer-scale contrast of material properties can be obtained.
In addition, it has been suggested that STM-SMM allows us to obtain an atomic-resolution microwave image using tunneling impedance~\cite{Lee2010,Reznik2014}.

Such SMMs have already been used for condensed matter research to evaluate the sheet resistances of metal oxide thin films~\cite{Park2001,Wang2005}, superconductors~\cite{Takeuchi1997,Steinheuer1997Dec} and combinatorial materials~\cite{Yoo2001,Okazaki2008}.
The value of the SMM will be further enhanced by cryogenic applications, and provide us with insights regarding the origin of various physical phenomena.
In fact, cryogenic SMM has revealed that intrinsic inhomogeneity is related to the metal-insulator transition~\cite{Hyun2002,Lai2010} and phase separation~\cite{HT2015pC,HT2015}. 
Additionally, SMM is expected to serve as a tool for locally studying inhomogeneous superconductors such as high-$T_c$ cuprate in the pseudogap phase~\cite{Lang2002,Kohsaka2007} and vortex matter~\cite{HT2012,Okada2012} because the microwave response directly reflects the low-energy quasiparticle dynamics~\cite{Maeda2005,HT2011}. 

In this paper, we present the design of a low-temperature-compatible STM-SMM utilizing a coaxial resonator probe.
We achieve stable operation at cryogenic temperature by designing a small and rigid microwave probe.
We also present the analysis of our SMM probe based on the simple lumped-element circuit model. 
Although we need a simulation method such as finite element method~\cite{Okazaki2007} for rigorous analysis, the lumped parameter model is useful for semi-quantitatively understanding the behavior of resonator probes.

\section{Apparatus}
\begin{figure}[tb]
	\begin{center}
		\includegraphics[width=0.9\hsize]{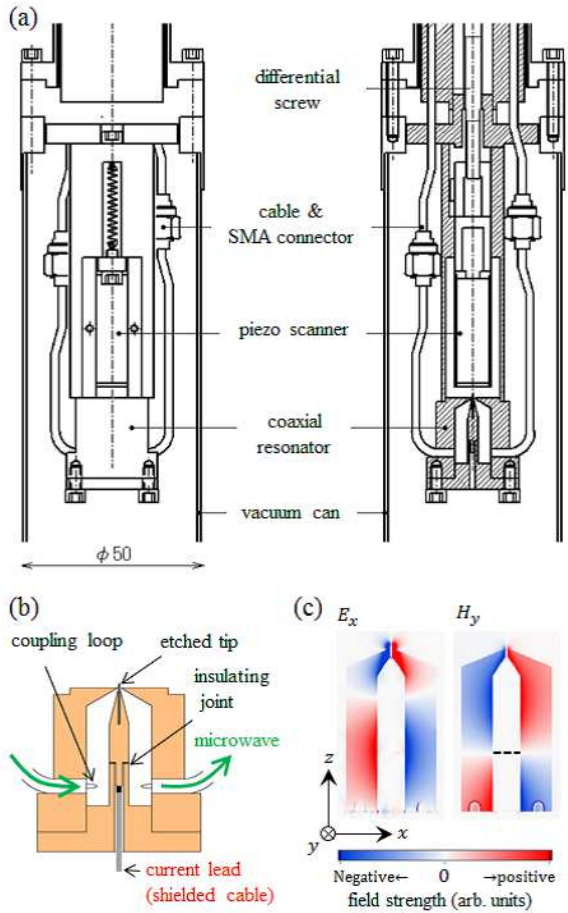}
	\end{center}
\caption{(a) Side (left) and cross-sectional (right) views of the SMM head. (b) Schematic of the coaxial resonator probe. (c) Simulation of the oscillating electric and magnetic fields inside the coaxial resonator probe. The red and blue color show the sign of the field direction. The dashed line in the right figure corresponds to the position of the insulating joint.}
\label{apparatus}
\end{figure}

Figure~\ref{apparatus}(a) shows our SMM head, which mainly comprise a resonator probe, a piezo scanner, and body parts.
The outer diameter at the thickest part is set to 50 mm to fit inside a superconducting magnet for the future experiment under magnetic field.
The body parts machined from oxygen-free copper (OFC) or brass form a coarse approaching mechanism, which is in direct contact with liquid helium at the upper flange to strongly cool the resonator probe.

The resonator probe (fig.~\ref{apparatus}(b)) was also made of OFC. 
It is similar to the often-used coaxial resonator~\cite{Wei1996}, except that the central conductor is divided into two parts.
There is a small hole at the tapered end of the upper part.
A sharpened metal tip is connected to this hole and protrudes out of the aperture of the outer conductor to simultaneously detect the tunneling current and microwave response.
We use a mechanically cut or electrochemically etched Pt-Ir wire as a tip, whose curvature at the end ($r_{\mathrm{tip}}$) is approximately 100 nm.
The upper part is glued to the bottom part with insulating varnish.
The shielded cable for the tunneling current detection is inserted through the hole of the bottom part, and then is soldered to the post of the upper part.
While this design separates the tunneling current and microwave circuits, it is not suitable for operation in the lowest transverse electromagnetic (TEM) $\lambda /4$ mode (where $\lambda$ is the wavelength) because of the considerable energy loss at the joint part. Instead, we use the second-lowest TEM $3\lambda /4$ mode at $f_0=2\pi/\omega_0=10.7\ \mathrm{GHz}$. 
Figure~\ref{apparatus}(c) shows the distribution of the electromagnetic field inside the resonator. In this mode, the electric and magnetic fields are almost confined to the radial and the angular direction, respectively.
The oscillating magnetic field has a node. 
The energy loss is significantly reduced by adjusting the joint position at this node. Consequently, we can maintain a high quality factor $Q$. 
The unloaded quality factor of the resonator is $Q_0=$1200-1300 at room temperature and $Q_0>$2000 below liquid nitrogen temperature.  

Semi-rigid coaxial cables (0.085 inches in diameter) are used for transmission and coupling to the resonator.
These cables are thermally anchored to the body part.
Sample and tip interchange is performed by disconnecting the SMA connectors.
After a sample is set, the vacuum can is sealed with indium wire; then, it is evacuated and helium is used as exchange gas.

\begin{figure}[tb]
	\begin{center}
		\includegraphics[width=0.9\hsize]{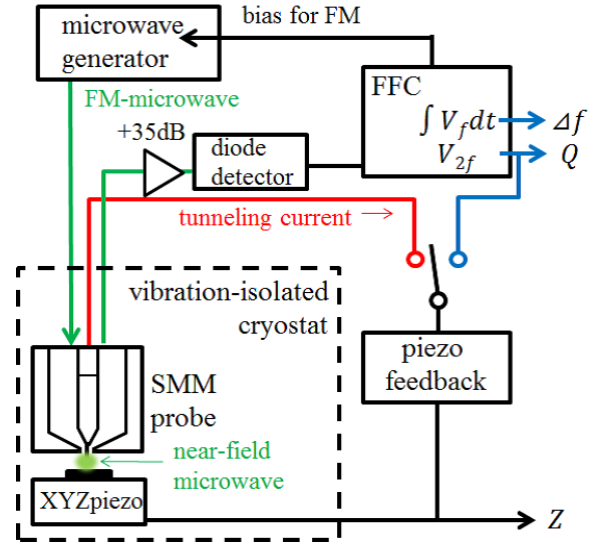}
	\end{center}
\caption{The schematic of the control circuit for SMM using FFC. Either the tunneling current or $Q$ is used as a feedback signal. }
\label{FFC}
\end{figure}

The measurement system comprises STM and microwave circuits.
A commercial STM controller  (RC4 and SC4 by SPECS Surface Nano Analysis GmbH) is used to control tunnel current feedback.
The quality factor, $Q$, and resonant frequency shift, $\Delta f$, are measured either by acquiring the transmission spectrum using  a PNA network analyzer or by the frequency feedback circuit (FFC)~\cite{Steinheuer1997July}.
In the latter method, a frequency-modulated (FM) microwave with a frequency expressed as $f_c+D\cos (2\pi f_m t)$ (where $f_c$, $f_m$, and $D$ are the carrier frequency, modulation rate, and frequency deviation, respectively) is used (fig.~\ref{FFC}).
The modulation is controlled by the DC-coupled external bias circuit.
When $f_c$ coincides with the resonant frequency of the coaxial probe, the transmitted power oscillates with a frequency of $2f_m$, which is converted into a voltage signal, $V_{2f}(\propto Q)$ by the diode detector (Agilent 8473C).
Otherwise, the component oscillating with $f_m$ ($V_f$) becomes large.
By feedback control minimizing $V_f$, $f_c$ is locked at $f_0$.
The advantage of the FFC method over the method that uses a PNA network analyzer is its high speed.
In addition, the FFC method allows us a tip-sample distance control not only by STM feedback (constant current mode) but also by keeping $Q$ constant (constant Q mode)~\cite{HT2015}. 
However, since the phase-locked loop inside the microwave source (HP83630A) should be open when the modulation is controlled by an external circuit, a drift of the source frequency arises.
A typical drift rate is approximately $10\ \mathrm{kHz/min}$, and its influence is corrected after data acquisition.
In this work, SMM is always operated at $f_0$. It should be mentioned that there are other studies that use slight off-resonant excitation to optimizing the measurement condition~\cite{Sardi2015,Gregory2016}.

\begin{figure}[tb]
	\begin{center}
		\includegraphics[width=1\hsize]{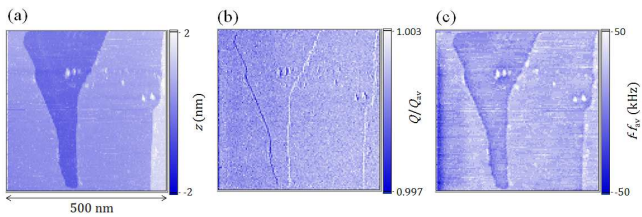}
	\end{center}
\caption{Simultaneously acquired (a) topography; (b) quality factor; and (c) frequency shift image of the cleaved surface of Bi$_2$Se$_3$ single crystal.}
\label{Bi2Se3}
\end{figure}

Figure.~\ref{Bi2Se3}(a)-(c) show example images taken by our SMM.
The sample is cleaved Bi$_2$Se$_3$ and a temperature is 77 K.
The $Q$ and $\Delta f$ images are raw images without any geometrical corrections, whereas a tilt correction is applied for topography.
These are obtained using the PNA network analyzer; hence, these are free from the frequency drift of the microwave source.
Therefore, these images directly reflect the performance of our SMM probe.
Since this sample is homogeneous, $Q$ and $\Delta f$ are constant in most regions. 
The resonant characteristics change only at the edges of the  terraces.
These changes are related to the change in the capacitance between the tip and sample, $C_x$, which will be discussed later.
Despite the long measurement time of approximately an hour, we obtain images without any drift of the microwave properties, indicating the high long-term stability of our SMM probe.

\section{Analysis}
\begin{figure}[tb]
	\begin{center}
		\includegraphics[width=0.9\hsize]{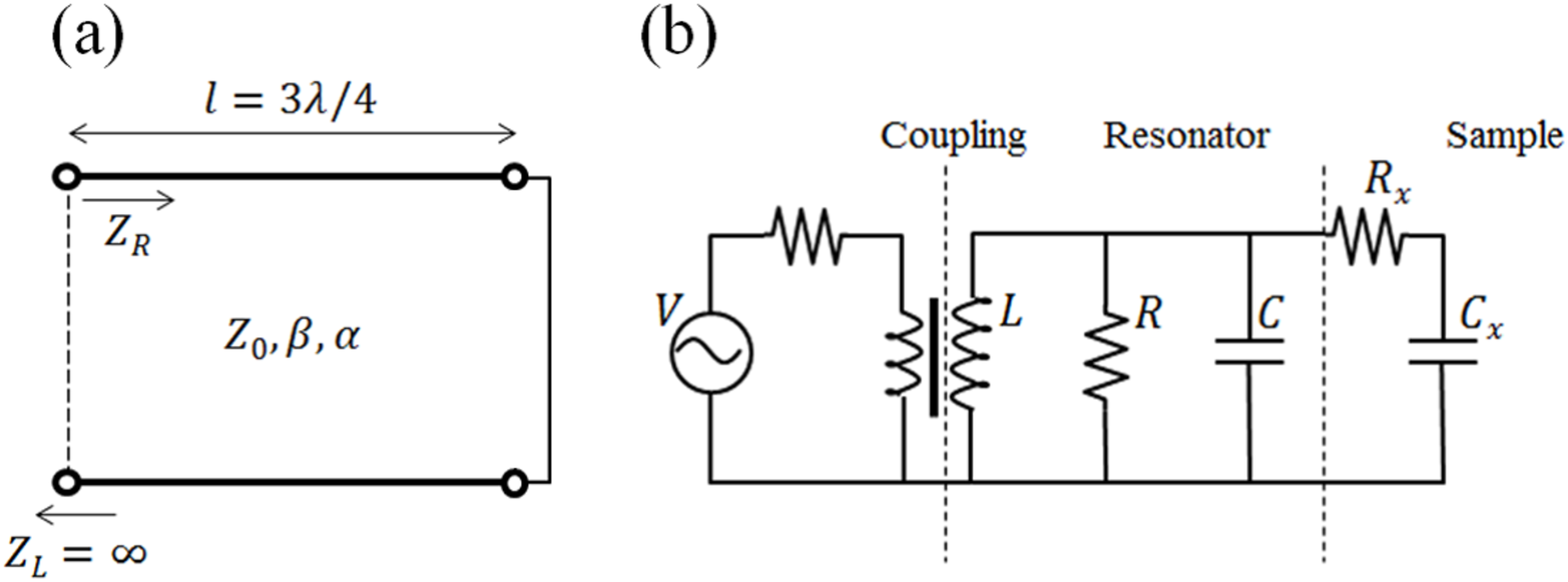}
	\end{center}
\caption{(a) The equivalent trasmission line and (b) the lumped element circuit of the coaxial resonator. }
\label{equivcircuit}
\end{figure}

To discuss the change in the resonant characteristics caused by samples having different conductivities, we first model the  resonator probe by a distributed constant circuit; then, we simplify it using a lumped parameter circuit.
The coaxial resonator is equivalent to a transmission line resonator described in fig.~\ref{equivcircuit}(a).
The left side corresponds to the tip end of the resonator.
$\gamma(=\alpha+j\beta)$ is a propagation constant, with $\alpha$ being an attenuation constant related to the loss in the transmission line and $\beta$ being a phase constant equal to $2\pi/\lambda$ for air.

In transmission line theory, the input impedance at the position $l$ away from the load impedance, $Z_{\mathrm{load}}$, is 
\begin{equation}
Z_{\mathrm{in}}=Z_{0}\frac{Z_{\mathrm{load}} + Z_0 \tanh \gamma l}{Z_{0}+Z_{\mathrm{load}} \tanh \gamma l}.
\end{equation}
Where $Z_0$ is the characteristic impedance of the transmission line, which is expressed as
\begin{equation}
Z_0=\sqrt{\frac{\mu}{\epsilon}}\frac{\ln b/a}{2\pi}.
\end{equation}
In our case, $Z_0 =70\ \mathrm{\Omega}$ using a ratio of the outer diameter to the inner one of $b/a\ =\ 3.3$. 
Since the resonator is closed at the right end, the impedance on the right-hand side of the dashed line in fig.~\ref{equivcircuit}(a) is
\begin{equation}
Z_{R}=Z_{0}\frac{\tanh\alpha l+j\tan \beta l}{1+j\tanh\alpha l\tan \beta l}.
\end{equation}
When the transmission line is lossless ($\alpha=0$), we obtain
\begin{equation}
Z_{R}=jZ_0\tan \beta l.
\end{equation}
When a sample is absent, the impedance of the tip end, $Z_L$, is $Z_L=\infty$. 
From the resonant condition met when $Z_R+Z_L=0$,
we obtain
\begin{equation}
l=\frac{\lambda}{4}(2n-1)\ \ \ \ (n=1,2,3\cdots)
\end{equation}
The TEM $3\lambda /4$ mode corresponds to the case of $n\ =\ 2$.

Next, we discuss the quality factor of the resonator by considering a finite transmission-line loss. When the loss is small ($\alpha l\ll 1$), $Z_R$ is expressed as
\begin{equation}
Z_{R}=Z_{0}\frac{\alpha l\cot \beta l +j}{\cot\beta l +j\alpha l}.\nonumber\\
\end{equation}
Near the resonant frequency $\omega_0$, $Z_R$ of a TEM $3\lambda /4$ resonator is
\begin{equation}
Z_{R}=Z_{0}\frac{1+3j\alpha l \pi\Delta \omega/2\omega_0 }{\alpha l+3j\pi\Delta \omega/2\omega_0}
\simeq\frac{Z_0}{\alpha l+3j\pi\Delta \omega/2\omega_0},
\end{equation}
where $\Delta\omega$ ($\Delta\omega/\omega_0 \ll 1$) is the deviation from the resonant frequency. This formation is equivalent to the input impedance of the lumped $RLC$ parallel circuit, $Z_{\mathrm{in}}=(R^{-1}+2j\Delta\omega C)^{-1}$,
whose quality factor and resonant frequency are $Q=\omega_0 RC$ and $\omega_0=1/\sqrt{LC}$, respectively. Therefore, our resonator probe can be simplified by the lumped element circuit shown in fig.~\ref{equivcircuit}(b).
The corresponding $R$ and $C$ components can be determined as
\begin{equation}
R=\frac{Z_0}{\alpha l}=\frac{2Z_0 Q}{\beta l}=\frac{2\times 70\times 2000}{3\pi/2}=6\times 10^4\ \Omega,
\end{equation}
\begin{equation}
C=\frac{3\pi}{4\omega_0 Z_0}=0.5\ \mathrm{pF}.
\end{equation}
Here we use $Q$ in the cryogenic environment, $Q_0=2000$.

\begin{figure}[tb]
	\begin{center}
		\includegraphics[width=1\hsize]{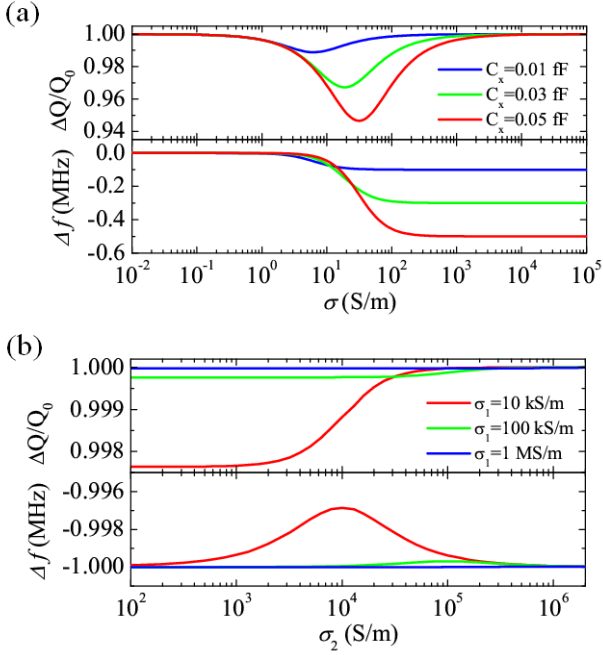}
	\end{center}
\caption{(a) $\sigma$-dependences of $Q$ and $\Delta f$ for different $C_x$ values. (b) $\sigma_2$-dependences of $Q$ and $\Delta f$. $C_x=0.1\ \mathrm{fF}$ was used for this calculation.}
\label{calculation}
\end{figure}

The tip-sample interaction is modeled by the series of the near-field impedance of a sample and the coupling capacitance, $Z_{\mathrm{load}}=R_x +1/j\omega C_x$. 
For conducting samples, $R_x$ is roughly approximated as $R_x=1/\sigma r_{\mathrm{tip}}$, where $\sigma$ is the DC conductivity. $C_x$ is calculated to be on the order of 0.01 fF by applying a parallel-plate approximation ($C_x \approx \epsilon_0 \pi r_{\mathrm{tip}}^2/h$, where $\epsilon_0$ is the dielectric permeability in vacuum and $h$ is the tip height from the surface). 
By a straightforward calculation, we obtain the changes in $Q$ and the resonant frequency as follows:
\begin{equation}
\frac{\Delta Q}{Q_0}\simeq-\frac{(\omega C_x R)(\omega C_x R_x)}{1+(\omega C_x R)(\omega C_x R_x)+(\omega C_x R_x)^2},
\end{equation}
\begin{equation}\label{fvsCx}
\frac{\Delta f}{f_0}\simeq-\frac{C_x}{2C}.
\end{equation}
$Q$ exhibits a minimal value when $\omega C_x R_x =1$ (fig.~\ref{calculation}(a)).
On the other hand, $\Delta f$ exhibits a monotonous $R_x$ dependence.

In applications to superconductors, we need to consider the contribution of the imaginary part of complex conductivity, $\tilde{\sigma}=\sigma_1+j\sigma_2$.
Since $\sigma_1$ and $\sigma_2$ are respectively proportional to the normal fluid and superfluid density, $\sigma_1\ll\sigma_2$ at a temperature sufficiently lower than $T_c$.
Figure~\ref{calculation}(b) shows the $\sigma_2$-dependence of the resonant characteristics at fixed $\sigma_1$.
$Q$ shows monotonous change against $\sigma_2$ with a maximum slope ($dQ/d\sigma_2$) at $\sigma_1=\sigma_2$ and approaches the value at the unloaded condition when $\sigma_1\ll\sigma_2$.
On the contrary, $\Delta f$ shows nonmonotonous change against $\sigma_2$.
The characteristic behavior is also observed at $\sigma_1=\sigma_2$ as a peak. 
It is noteworthy that one cannot distinguish superconducting and non-superconducting phases only by a frequency measurement because a difference of less than $<1\ \mathrm{kHz}$ in $\Delta f$ is beyond the frequency resolution of our setup.
On the contrary, a 0.1 \%-difference in $Q$ between the superconducting and metallic limits is detectable.

\begin{figure}[tb]
	\begin{center}
		\includegraphics[width=1\hsize]{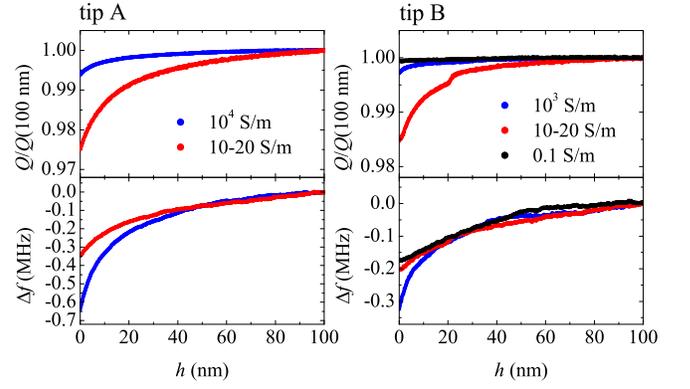}
	\end{center}
\caption{The tip height-dependences of $Q$ and $\Delta f$ measured for silicon wafer samples with different conductivities. }
\label{Siwafers}
\end{figure}

To examine the model's usefulness, we measure the near-field response of bulk silicon wafers with different conductivities at room temperature. 
Figure~\ref{Siwafers} shows the $h$-dependencies of the resonant characteristics.
Different tips (A and B) are used in the left and right figures.
$Q$ is normalized by the values at $h=100\ \mathrm{nm}$.
Both $Q$ and $\Delta f$ monotonically decrease as the tip approaches the sample because of the increase of $C_x$.
When we compare the data with the value at $h=0$, we find a remarkable difference between the semiconducting samples ($\sigma=10$-$20\ \mathrm{S/m}$) and the metallic ones ($\sigma=10^3\ \mathrm{S/m}$, $10^4\ \mathrm{S/m}$) ; the change in $Q$ is larger for the semiconducting samples, while the change in $\Delta f$ is larger for the metallic samples. 
For the insulating sample ($\sigma=0.1\ \mathrm{S/m}$), both $Q$ and $\Delta f$ exhibit their smallest $h$-dependencies.
The lumped element model predicts that the change in $C_x$ will be the dominant factor affecting $\Delta f$ (Eq.~\ref{fvsCx}).
$\Delta f$ below $h=100\ \mathrm{nm}$ corresponds to the 0.01-0.1 fF change in $C_x$, which is of the same order as the calculated value with a simple parallel-plate approximation.

The lumped element model semi-qualitatively explains the behavior of the SMM probe.
However, it is too simplified for quantitative discussion.
We assume that the electromagnetic field is tightly localized around the tip with a decaying length comparable to $r_{\mathrm{tip}}$.
In reality, the field decays in a power-law manner~\cite{Imtiaz2006}. 
A more rigorous model requires modifying the expression for the load impedance, for which the tip geometry and field distribution are considered~\cite{Okazaki2007}.

\section{Imaging}

To evaluate the performance of SMM at cryogenic temperature, we measure the spatial distribution of the microwave response of a single crystal of the iron chalcogenide superconductor Fe(Se,Te) with a PbO structure~\cite{Fang2008}. 
The sample is grown by the Bridgeman method from starting materials with a nominal Se:Te ratio of 0.4:0.6.
Although the annealed crystal exhibits perfect shielding below $T_c=14\ \mathrm{K}$~\cite{HT2011}, the unannealed crystal separates into superconducting and non-superconducting phases having different chemical composition~\cite{Speller2011,Speller2012,Prokes2015}.
The origin of the phase separation is considered to be a large difference in the Fe-Se and Fe-Te bonding lengths~\cite{Joseph2010}.

\begin{figure}[bt]
	\begin{center}
		\includegraphics[width=1\hsize]{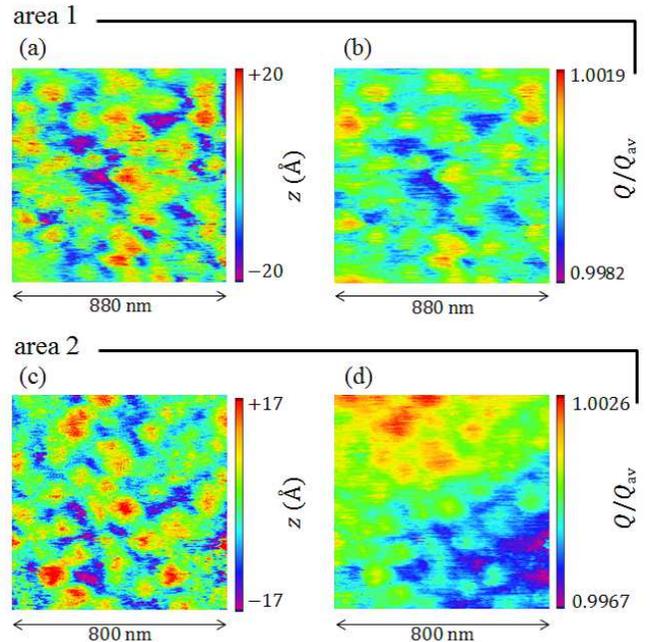}
	\end{center}
\caption{(a)-(d) Topographies and $Q$ images of Fe(Se,Te) acquired in a homogeneous area (area 1) and an inhomogeneous area (area 2). }
\label{FeSeTe_CC}
\end{figure}

Figure~\ref{FeSeTe_CC} shows the surface topography and spatial dependence of $Q$ for  Fe(Se,Te), acquired in the constant current mode.
The most typical images are figs.~\ref{FeSeTe_CC}(a) and \ref{FeSeTe_CC}(b). 
The surface roughness is less than 5 nm.
The changes in the areas where topographical changes are observed are attributed to
the abrupt change of $C_x$.
As discussed above, near-field microwave measurement is sensitive not only to $R_x$ but also to $C_x$.
In fact, the sharp changes at the edges of the terraces in figs.~\ref{Bi2Se3}(b) and \ref{Bi2Se3}(c) are also attributed to the change in $C_x$.
As $C_x$ strongly depends upon the tip-sample distance and geometry, even a 1-nm step affects the microwave images.
Although this is often problematic, one can distinguish whether the contrast in images is related to sample properties or is only a geometrical artifact by carefully examining the topography and microwave image.
If we find that the contrast does not correlate with topographic change in the microwave images, we can conclude that it is caused by electric inhomogeneity. 
Figures.~\ref{FeSeTe_CC}(c) and \ref{FeSeTe_CC}(d) are images of the other region. In addition to the $C_x$-induced contrast, we can see the changes in the microwave response that are not correlated with topography; the contrast gradually changes from the upper left to lower right. 
This change is considered to be related to the inhomogeneous $R_x$.

The more effective method for separating the topographical and electrical contrast requires slight modification of the measurement system from the constant current mode.
In this method, we select $Q$ as a feedback signal instead of the tunneling current. 
The tunneling current is monitored only so that the tip does not contact the sample.
$h$ changes depending upon the local material property while $Q$ is kept constant.
Since $\Delta f$ largely depends on $h$, it exhibits significant changes only when the tip crosses a boundary between regions with different conductivities. 
As a result, a qualitative image is obtained.
The advantage of this scanning mode is that the topographical information is largely eliminated in the obtained image.
We can avoid the influence of the fluctuation of $C_x$ by setting $h$ higher than the constant current mode (typically $h=10$-$20\ \mathrm{nm}$).  

\begin{figure}[bt]
	\begin{center}
		\includegraphics[width=1\hsize]{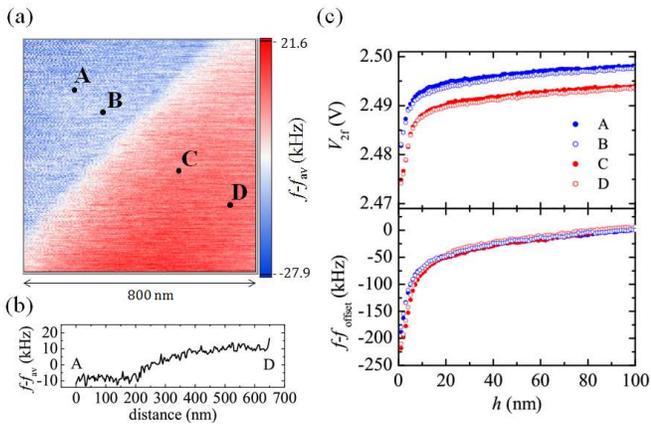}
	\end{center}
\caption{(a) The frequency image acquired in the constant $Q$ mode for the same region as the area 2 in fig.~\ref{FeSeTe_CC}. (b) The linecut between the positions A and D in (a). (c) The tip height-depedences of $V_{\mathrm{2f}}(\propto Q)$ and the resonant frequency at the positions A-D. The offset frequency, $f_{\mathrm{offset}}=10.720974\ \mathrm{GHz}$, is subtracted from the data for clarity.}
\label{FeSeTe_CQ}
\end{figure}

Figure~\ref{FeSeTe_CQ}(a) is the frequency image acquired in the constant $Q$ mode for the same region as area 2 in fig.~\ref{FeSeTe_CC}.
The topographic contrast that was present in the constant current image (fig.~\ref{FeSeTe_CC}(d)) has disappeared.
As a result, we can find the boundary between two different phases. 
Fig.~\ref{FeSeTe_CQ}(c) shows the $h$ dependencies of the microwave response at positions corresponding to A-D in fig.~\ref{FeSeTe_CQ}(a).
The sharp change below $h=5\ \mathrm{nm}$ is caused by a polluted layer on the sample surface.
What is important is the behavior above $h=5\ \mathrm{nm}$. 
$Q$ in the blue region (positions A and B) is higher than that in the red region (positions C and D), and its difference is observed even at $h=100\ \mathrm{nm}$.
Since the length scale of  the near-field microwave is approximately $r_{\mathrm{tip}}+h$ when the tip is at height $h$~\cite{Imtiaz2006}, these data indicate that the length scale of electric inhomogeneity is much larger than 100 nm.
On the contrary, the $h$ dependencies of the frequency shift does not exhibit a significant difference between the two regions.
The change solely observed in $Q$ suggests that the contrast is related to the difference in $\sigma_2$.
As shown in fig.~\ref{calculation}(b), high $Q$ is observed in the superconductive region, while the superconductivity hardly affects $\Delta f$.
Therefore, in fig.~\ref{FeSeTe_CQ}(a), the upper-left and lower-right regions correspond to the superconducting and non-superconducting regions, respectively.

As the tip crosses the boundary between different phases, the change in the microwave response occurs within a width of 200 nm (fig.~\ref{FeSeTe_CQ}(b)). If we assume that the sample has a well-defined boundary and that the electrical property is homogeneous in each region, this indicates that the spatial resolution is no worse than 200 nm. This value is consistent with the curvature of the tip.

The use of sharper tips is indispensable for further improvement of the spatial resolution.
However, it is expected that spatial resolution does not improve linearly with $r_{\mathrm{tip}}$ decreases.
At small $r_{\mathrm{tip}}$ values, the near-field response is very small because of the weak coupling between the tip and sample.
In such a situation, we have to consider the proximity effect, i.e., the contribution from an additional capacitive component that arises from a finite aspect ratio of the tip.

The combination of CC and CQ scanning modes is applicable also for distinguishing semiconducting phase from metallic phase.
Our previous studies have revealed web-like mesoscopic phase separation in KFe$_x$Se$_2$ at room temperature~\cite{HT2015pC,HT2015}.

\section{Summary}
We have developed the low-temperature compatible STM-SMM.
The modified coaxial resonator probe allows STM operation without sacrificing the high $Q$ factor.
The behavior of the SMM probe was described by the lumped element circuit, which was confirmed by the near-field response to silicon wafers having different conductivities. 
We also demonstrated that STM-SMM can be used for the study of inhomogeneous superconductors using two scanning modes.
The spatial resolution is approximately 200nm at this time, which is as high as that in previous reports~\cite{Imtiaz2007,Machida2009}.

A challenge for the future is to combine microwave measurement with local tunneling spectroscopy~\cite{Machida2009}.
Since both the local density of states and microwave conductivity are important for understanding the nature of quasiparticles in the superconductor, STM-SMM will be a useful tool for studying nanoscale inhomogeneity and the vortex state in superconductors.

The authors thank for Yusuke Yasutake and Tetsuo Hanaguri for technical assistance, and Susumu Fukatsu for providing us silicon wafers.
This work has been supported by a Grant-in-Aid for Scientific Research(A) (Grants. No. 23244070) from the Ministry of Education, Culture, Sports, Science, and Technology of Japan. H. Takahashi also thanks the Japan Society for the Promotion of Science for financial support.

\end{document}